\documentclass[aps,pre,twocolumn,showpacs,superscriptaddress]{revtex4}
\usepackage{epsfig}
\usepackage{times}

\begin{document}

\title{Alliance formation with exclusion in the spatial public goods game}

\author{Attila Szolnoki}
\email{szolnoki@mfa.kfki.hu}
\affiliation{Institute of Technical Physics and Materials Science, Centre for Energy Research, Hungarian Academy of Sciences, P.O. Box 49, H-1525 Budapest, Hungary}

\author{Xiaojie Chen}
\email{xiaojiechen@uestc.edu.cn}
\affiliation{School of Mathematical Sciences, University of Electronic Science and Technology of China, Chengdu 611731, China}

\begin{abstract}
Detecting defection and alarming partners about the possible danger could be essential to avoid being exploited. This act, however, may require a huge individual effort from those who take this job, hence such strategy seems to be unfavorable. But structured populations can provide an opportunity where largely unselfish excluder strategy can form an effective alliance with other cooperative strategies hence they can sweep out defection. Interestingly, this alliance is functioning even at extremely large cost of exclusion where the sole application of exclusion strategy would be harmful otherwise. These results may explain why the emergence of extreme selfless behavior is not necessarily against individual selection, but could be the result of an evolutionary process.
\end{abstract}

\pacs{89.75.Fb, 87.23.Kg}

\maketitle

\section{Introduction}
It is always disappointing to realize exploitation by others in a joint venture where participants decide independently whether to contribute to a common pool. The core of social dilemmas origins from the fact that players in general have no preliminary information about how partners will behave \cite{nowak_11, mestertong_01, sigmund_10}. This problem can be addressed in several ways, like by recording previous acts via a reputation-like tag, by monitoring and rewarding cooperators or punishing defector who become less successful in this way \cite{fehr_n04, brandt_prsb03, milinski_n02, szolnoki_epl10, fu_pre08b, masuda_prsb07b, laird_ijbc12, traulsen_plosone07, szolnoki_njp12, zhang_h_amc16, capraro_jdm16, javarone_epl16, wang_z_jtb14, cuesta_jtb08, wu_y_srep17,rezaei_ei09,  gao_l_srep15, szolnoki_prsb15, wang_xw_pa17}. But every kind of approach requires an additional individual effort from the player who pays extra attention to check others. This possibility transforms the dilemma onto a new level, where the question is who bears the extra cost of monitoring and punishing defection \cite{panchanathan_n04,cui_pb_jtb14, ye_h_srep16}? Interestingly, this so-called second-order free-riders dilemma may be solved automatically in structured populations where players have limited number and practically stable links which allow punishing players to separate from pure cooperators who can be considered as second-order free riders \cite{helbing_ploscb10}. When these strategies form isolated groups in the sea of defectors then the advantage of punishing strategy reveals and pure cooperators diminish, which results in a higher average cooperator level.

Punishing others, however, is not necessarily an attractive strategy for every cooperator and the application of punishment will reduce the average payoff in the population \cite{fehr_n02, de-silva_jee10}. An alternative way to avoid being exploited is if a player monitors defection and alarm all other cooperator players in the group about the possible danger. The mentioned player hence can exclude defectors from the joint venture for the benefit of all others. Some previous works have already studied the possible positive effect of social exclusion in well-mixed populations \cite{sasaki_prsb13,li_k_pre15,li_k_epl16,liu_lj_srep17}, but its consequences in structured populations remained unexplored. The goal of our present work is to reveal how exclusion strategy may influence the evolution of cooperation in a system where network reciprocity already establishes a supporting environment for cooperation. 

For this reason we extend a previously studied model where beside pure cooperators punishing players also compete with defectors for space in a structured population \cite{helbing_njp10}. In our present model we add a new strategy, which is called as excluder, who undertakes the extra effort of monitoring and excluding defectors from the joint venture. Our principal goal is to identify the limit of excluder's cost until this strategy is able to survive and serves the community. We will show that the most effective way to fight against defection is when exclusion strategy forms a defensive alliance with other cooperative strategies and they support each other mutually to sweep out defectors. First, however, we proceed with presenting the details of the extended spatial model. 

\section{Public goods game with exclusion}
\label{def}

We study an evolutionary public goods game (PGG) on the square lattice where players can choose from four different strategies. The competing strategies are $D$ defectors who do not contribute to the common pool, but only enjoy its positive consequence, $C$ cooperators who contributes $c=1$ to the joint venture, but do not bear the extra cost of punishment or exclusion, $P$ punishers who do not only contribute to the pool, but also punish defectors at the expense of an extra cost, and finally $E$ excluders who beside contributing to the pool also monitor defectors in the group and alarm all other group members about the possible danger. Due to this alarming a defector will be excluded from the common game and returns empty handed, while other group members can enjoy the benefit of common acts. Perhaps it is worth noting that social exclusion maintains only if $E$ players are present in the groups, otherwise, when they are absent, defector player can exploit group efforts again. Evidently, monitoring and alarming others require an extra effort from an $E$ player which is considered via an extra cost. We should also stress that the cost of punishment and social exclusion should not necessarily be equal, hence we can study the specific cases when an excluder may bear an extremely high cost.

Denoting the number of cooperators, defectors, punishers, and excluders among neighbors in the group by $N_D, N_C, N_P$, and $N_E$ respectively, the payoff of the focal player is:
\begin{eqnarray}
\Pi_D\,\, &=& \delta (N_E) \frac{r (N_C+N_P)}{G} - \beta \frac{N_P}{G-1} \\
\Pi_C\,\, &=& \delta (N_E) \frac{r (n_C+N_P+1)}{G} + (1-\delta (N_E)) r- 1\,\,\,\,\,\,\,\,\,\,\\
\Pi_P\,\, &=& \Pi_C - \gamma \frac{N_D}{G-1}\\
\Pi_E &=& r - 1 - \epsilon \frac{N_D}{G-1}\,\,,
\label{payoff}
\end{eqnarray}
where the Kronecker-delta function is $\delta(x)=1$ if $x=0$, otherwise $\delta(x)=0$. By using this delta function we can handle the situation when a defector player cannot gain anything from the group venture due to the presence of an excluder. The other parameters are idential to those we used in a simplified model \cite{helbing_njp10}. Namely, $r$ denotes the synergy factor, $\beta$ is the maximum value of fine of a defector when it is surrounded by punisher players exclusively. Parameter $\gamma$ denotes the maximum value of punishment cost that should bear by a punisher when it forms a group with four other defectors. Last, $\epsilon$ describes the additional cost of exclusion strategy which should always be considered in the presence of defectors. Equation~(4) indicates that our model considers peer exclusion. Consequently, a player who considers exclusion strategy pays an extra cost that is proportional to the number of defectors in the group. It also means that an $E$ player should not bear this cost in the absence of defectors. An alternative way to include social exclusion would be the so-called pool exclusion when $E$ strategy has to pay a fixed, but permanent extra cost independently of how many defectors are in the group. 

In the simplest case of a structured population the players are arranged on a square lattice with periodic boundary conditions where $L^2$ players are assigned to overlapping groups of size $G=5$ such that everyone is connected to its $G-1=4$ nearest neighbours \cite{perc_jrsi13}. Accordingly, each individual $i$ belongs to $g=1,\ldots,G$ different groups and the total payoff is the sum of all the payoffs $\Pi_{i}$ acquired in each individual group. We should stress, however, that our most important observations are robust and remain valid if we apply other interaction topologies, as it will be illustrated in the next section. The only essential criterion is to have limited number of neighbors who are fixed, at least for reasonable time of interaction comparing to the strategy update timescale \cite{pacheco_prl06, gomez-gardenes_c11, perc_bs10,pei_zh_njp17}.

To model the selection process during the evolution we apply an imitation dynamics \cite{schlag_jet98, szabo_pr07}. More precisely, during an elementary Monte Carlo step we select a player $x$ and one of its neighbor $y$ randomly. The total payoff $\Pi_{s_y}$ and $\Pi_{s_x}$ are calculated for both players. After player $y$ adopts the strategy from player $x$ with a probability given by the Fermi function $w(s_x \to s_y)=1/\{1+\exp[(\Pi_{s_y}-\Pi_{s_x}) /K]\}$, where $K=0.5$ quantifies the uncertainty by strategy adoptions \cite{szolnoki_pre09c}. This formula helps to avoid trapped artificial states \cite{sysiaho_epjb05} and ensures that better performing players are readily adopted, although it is not impossible to adopt the strategy of a player performing worse. Each full Monte Carlo step ($MCS$) gives every player a chance to change its strategy once on average.

\begin{figure}
\centerline{\epsfig{file=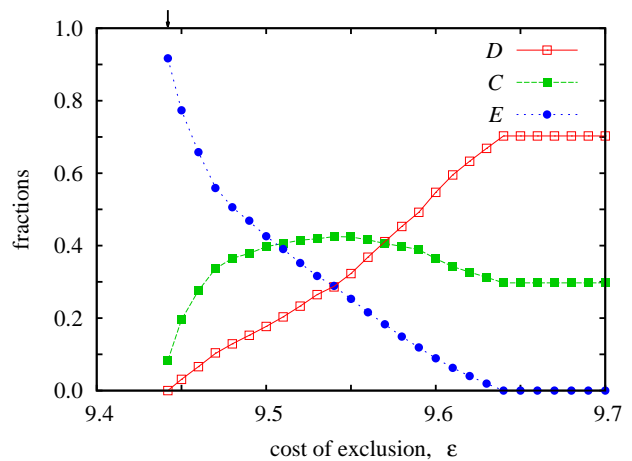,width=9.0cm}}
\caption{\label{hCP} (color online) Fraction of surviving strategies in dependence of the cost of exclusion at $r=3.8, \beta=0.2$, and $\gamma=0.6$. The high cost of punishment prevents strategy $P$ to survive and the remaining three strategies coexist in an intermediate interval of exclusion cost. If $\epsilon$ is too high then we get back the outcome of the classic two-strategy model where $D$ and $C$ coexist for high synergy factor values. Below a threshold value of $\epsilon$, marked by an arrow, defectors die out and the system terminates into a defector-free state. To avoid accidental die out of strategies due to fluctuation and to obtain the proper stationary values of cyclically dominating strategies we needed to use large system sizes (at least $L=1200$ and $1600$).}
\end{figure}

Note that in the absence of exclusion our model becomes equivalent with the previously studied PGG model with peer-punishment \cite{helbing_ploscb10,szolnoki_pre11b,helbing_njp10}. It is also worth stressing that the relations of $C, P$, and $E$ strategies  are neutral in the absence of defectors because neither $P$ nor $E$ has to bear an extra cost. In the latter case the trajectory of evolution becomes equivalent to a voter-model-like dynamics \cite{cox_ap86, ben-naim_pre96,dornic_prl01}. 
 
\section{Results}
\label{results}

Previous studies in structured populations explored that the value of synergy factor could divide the parameter space into two significantly different regions \cite{brandt_prsb03, helbing_ploscb10, helbing_njp10}. More precisely, if synergy factor is large enough then network reciprocity alone is capable to maintain cooperation and cooperators coexist with defectors. (This critical value is $r\approx3.744$ for square lattice at $K=0.5$ \cite{szolnoki_pre09c,perc_epj17}.) 
In the latter case, when network reciprocity is functioning, the final outcome depends practically on the cost of punishing strategy. If the cost of punishment is moderate then punishing players can crowed out pure cooperators via an indirect territorial competition because the former strategy is more effective against defectors \cite{helbing_ploscb10}. In the opposite case, when the cost of punishment is too large then we get back the outcome of the simplified two-strategy model where pure cooperators can coexist with defectors because $C$ players form compact islands in the sea of defectors and support each other mutually. The other conceptually different parameter region is when the synergy factor $r$ is too low and network reciprocity cannot really help to maintain cooperation. In the latter case neither $C$ nor $P$ strategy is able to coexist with defectors. As a result, $D$ prevails, or becomes extinct only if the punishment fine $\beta$ is large enough. The critical threshold of $\beta$ depends on the cost of punishment, as it is illustrated in Fig.~1(a) in Ref.~\cite{helbing_ploscb10}. 

\begin{figure}
\centerline{\epsfig{file=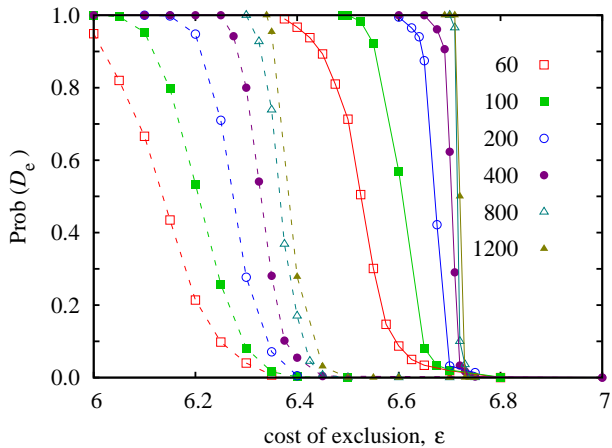,width=9.0cm}}
\caption{\label{D_E} (color online) Fixation probability to full-$E$ state at $r=3.8$ in dependence of cost of exclusion in the two-strategy ($D-E$) model. The alternative outcome of evolution is always the full-$D$ state, hence the plotted probability determines how likely defection dies out. This quantity shows a serious finite-size effect (linear sizes of systems are denoted by the legend). Furthermore, the character of initial state will also determine the speed of convergence to the large-size limit. In particular, symbols with dashed lines show the results when system was launched from a random initial state, while symbols with solid lines denote the results when evolution was launched from a patch-like pattern. Simulations were averaged over 10000 independent runs for small sizes, while 100 runs were used for the largest system sizes.}
\end{figure}

Based on these observations we can distinguish three conceptually different parameter regions where qualitatively different behaviors are expected if we add a new $E$ strategy of social exclusion. In the following we will focus on these three parameter regions and clarify whether the presence of $E$ strategy will change the evolutionary outcome.

\subsection{Behaviors in the high $r$ -- high $\gamma$ region}

We first start our presentation with the high synergy factor -- high punishment cost region where the simplified model would predict the coexistence of $C+D$ strategies. Note that by adding strategy $E$ into the system the number of parameters is also increased. Therefore, for a more complete view, we present the evolutionary outcomes in dependence of the exclusion cost $\epsilon$ for a representative value of $\gamma$. Figure~\ref{hCP} shows that excluders die out if the cost of exclusion is too high and cooperators and defectors coexist as in the classic two-strategy model. By decreasing $\epsilon$, however, we can observe a stable coexistence of $D, C$, and $E$ strategies, where the fraction of defectors decreases gradually with $\epsilon$. This kind of solution is a characteristic behavior in structured populations and it cannot be observed in well-mixed populations \cite{sasaki_prsb13, li_k_pre15}. Below a threshold value of $\epsilon$, marked by an arrow in the plot, defectors die out and only cooperator strategies remain. As already noted, after the extinction of the last defector a neutral drift starts among the surviving cooperating strategies in agreement with Eqs.~(2-4).  

\begin{figure}
\centerline{\epsfig{file=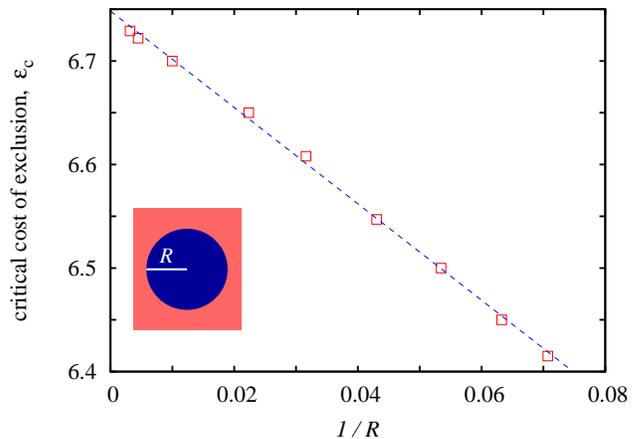,width=9.0cm}}
\caption{\label{critical} (color online) Curvature dependence of critical cost of exclusion when $E$ competes with $D$ at $r=3.8$. Initially $E$ players, marked by blue (dark grey), are arranged into a circle-shape island with radius $R$ in the sea of defectors, marked by red (light grey). If $\epsilon < \epsilon_c$ the island grows, otherwise shrinks. This behavior shows a sharp change at a fixed $R$. The border values of this change were determined from 100 independent runs for each $R$ value.}
\end{figure}

We would like to emphasize that the value of critical cost of exclusion $\epsilon_c=9.442$ is extremely high. It means that defector behavior can be crowded out totally from the population by means of exclusion strategy even if the latter players have to bear such an irrationally high extra cost. 

To evaluate this threshold value properly we now consider a simplified model in which only $D$ and $E$ strategies compete for space. In the latter case the system will always terminate into a full-$E$ or a full-$D$ state. Interestingly, this evolution is stochastic and both destinations can be observed by using identical parameter values. This serious finite-size effect is demonstrated in Fig.~\ref{D_E} where we plotted the probability to terminates to the full-$E$ state (or alternatively, the probability of strategy $D$ becomes extinct is shown) in dependence of exclusion cost for different system sizes. Beside the random initial state we have also used a so-called patch-like initial state where competing strategies are forming homogeneous domains from the beginning. (For clarity, an illustration of a prepared, patch-like starting state for 4 strategies can be seen in the inset of Fig.~\ref{lr}.) The consequences of the random initial state are plotted by dashed curves while the results of patch-like starting state are marked by solid lines. The first interesting observation is that the final destination of evolution depends sensitively on the total size of population. For example, at $\epsilon \approx 6.15$ both full-$E$ and full-$D$ states are equally likely for $L=60$, but using the same cost the system always arrives to the full-$E$ state if the system size exceeds $L=200$. As we increase the system size the related curves become steeper and final outcome becomes less ambiguous. Simultaneously, the critical cost shifts toward higher values. Interestingly, the convergence to the large-size limit behavior is much faster if we apply patch-like initial state to launch the evolution.

\begin{figure}
\centerline{\epsfig{file=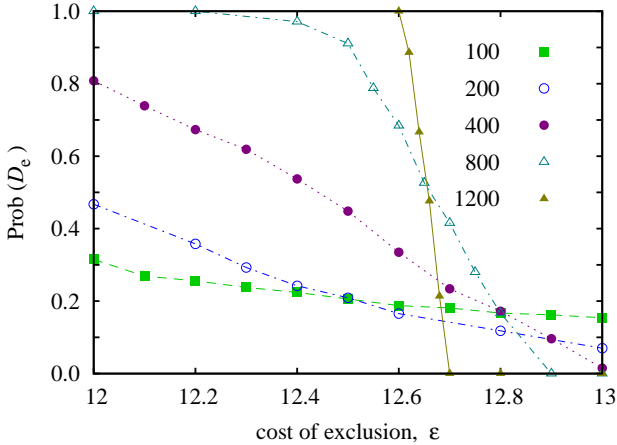,width=9.0cm}}
\caption{\label{lCP} (color online) The probability of defector's extinction when all 4 strategies are present in the initial state in a patch-like distribution illustrated in the inset of Fig.~\ref{lr}. The alternative outcome of the evolution is when strategy $E$ dies out and the system terminates to the well-known $D+P$ state. Parameters are $r=3.8, \beta=0.2$, and $\gamma=0.01$. The system sizes are marked in the legend.}
\end{figure}

This finite-size effect can be explained by the fact that the emerging size of a cluster containing $E$-players is strongly limited by the whole system size. If the latter is too small then only small $E$ islands can emerge after initial transient period and their possible growth is determined by the curvature of the interface separating competing $E$ and $D$ strategies. This effect can be measured systematically if we monitor the growth of $E$ islands with different sizes in the sea of defector players. For clarity, the initial state is shown in the inset of Fig.~\ref{critical}. If we launch the evolution from this state then the island will grow or shrink depending on the value of $\epsilon$. This critical $\epsilon$ value increases as we decrease the curvature of separating wall and converges to $\epsilon_c = 6.749$ in the limit when strategies compete for space along a straight domain wall. The latter can only emerge spontaneously if the system size is large enough, which explains the critical threshold value we obtained in Fig.~\ref{D_E}. To close our discussion about the origin of finite-size effects we should note that it is always a potential danger in multi-strategy spatial systems which is frequently ignored by agent-based simulations. Beside inappropriate initial states the system size may also limit the largest emerging characteristic length of patterns, which can also be a source of misleading conclusions. These problems can only be avoided by applying systematic finite-size analysis \cite{szolnoki_pre09d, lutz_g17, szolnoki_njp16}.

Interestingly, the critical cost value for the two-strategy system is significantly smaller than the threshold value ($\epsilon=9.442$) we obtained in Fig.~\ref{hCP} when all possible strategies were present during the evolution. The difference of critical cost values highlights that there is a kind of synergy between $C$ and $E$ strategies and they can sweep out all defectors together, but strategy $E$ alone would not be able to fight against $D$ efficiently at such a high cost. This synergy is especially interesting because in general pure cooperators are believed as "second-order free riders" who do not bear the extra cost of punishment or exclusion, hence they just utilize the positive consequence of the latter acts \cite{panchanathan_n04, fehr_n04}. Our observation suggests that the presence of "less cooperative" $C$ strategy could be vital, because without them the largely unselfish excluder strategy would not function properly. This alliance could be effective even if strategy $E$ has to bear an unrealistically high individual cost and may explain why we can find examples of selfless behavior abound in human societies \cite{nowak_sciam12}. As we will illustrate later the formation of conceptually similar alliance  with punisher strategy is also possible and this effect utilizes strongly the limited interactions of players provided by structured population. 

\begin{figure}
\centerline{\epsfig{file=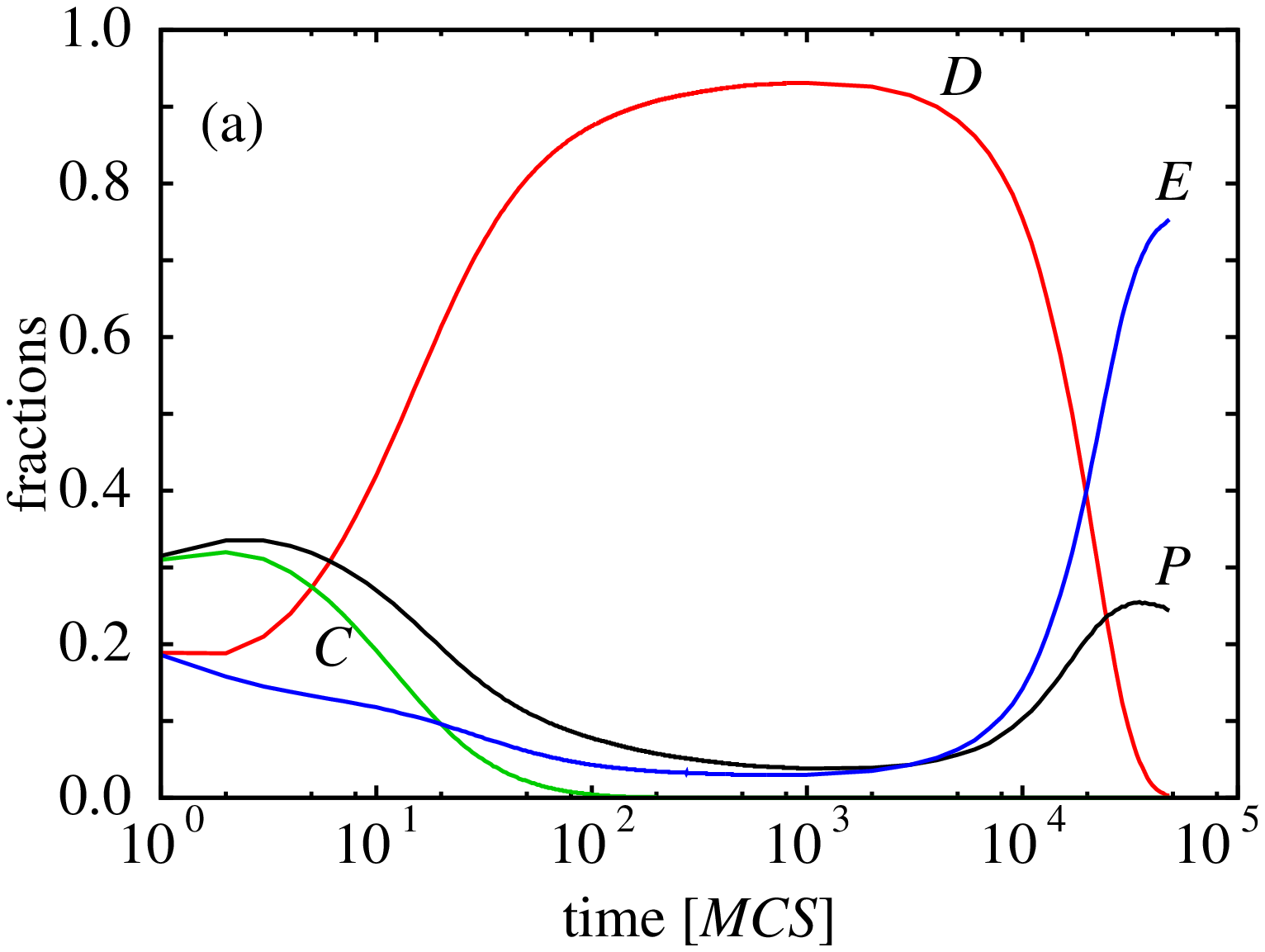,width=4.4cm}\epsfig{file=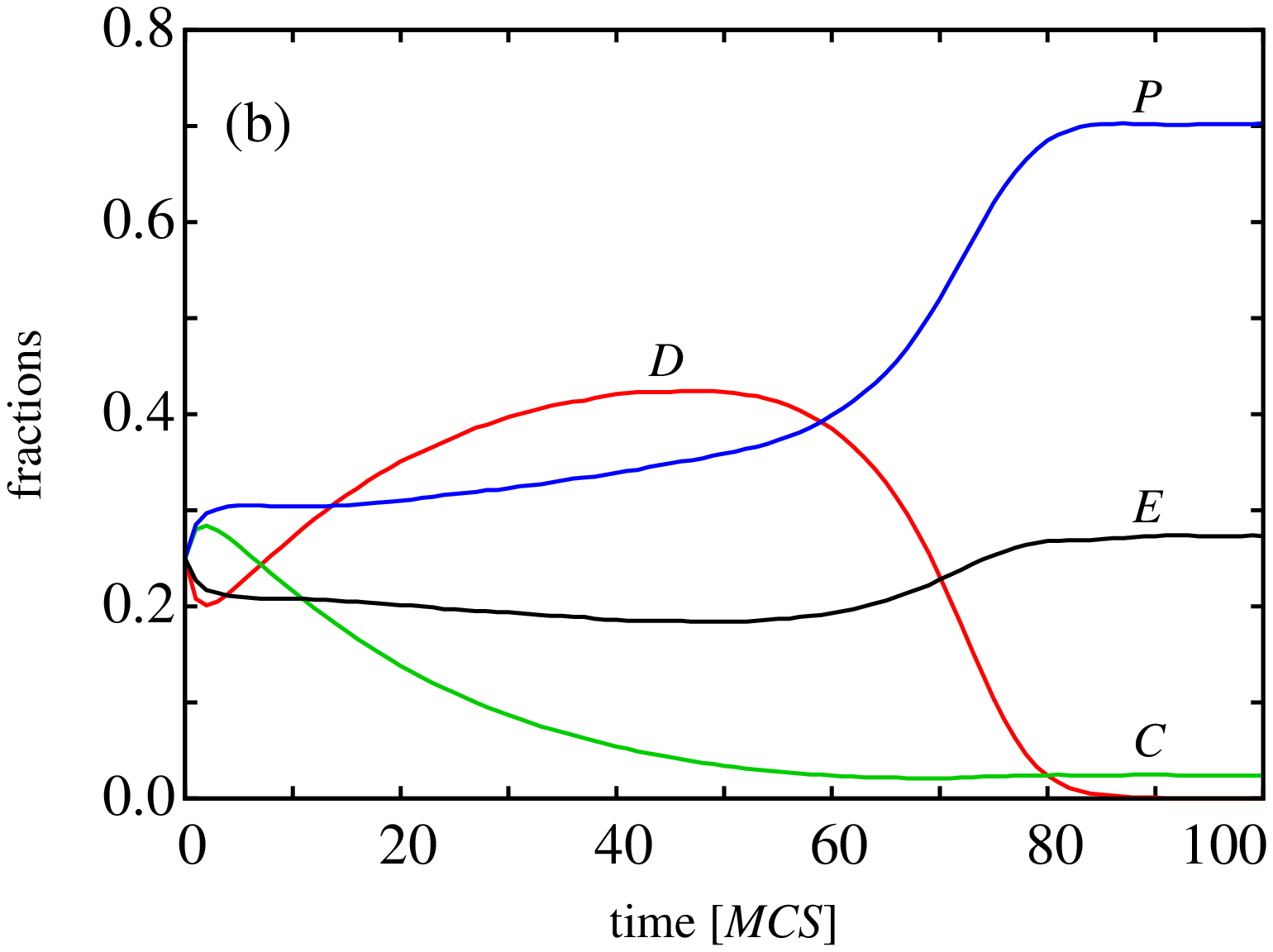,width=3.96cm}}
\caption{\label{monitor} (color online) Panel~(a): The time evolution of the frequencies of competing strategies starting from a random initial state on square lattice. Parameters are $r=2, \beta=0.6, \gamma=0.01, \epsilon=2.56$, and $L=6000$. The time courses suggest that cooperating strategies are week alone against defection and they can only survive if they form an effective alliance. When such an island containing $E$ and $P$ players emerge then they can sweep out all defectors gradually. To present the initial evolution clearly logarithmic time scale was used. Panel~(b) illustrates that similar behavior can be observed in random graph, but the whole evolution is much faster due to short-cut links between players. Parameter values are $r=1.44, \beta=0.7, \gamma=0.05$, and $\epsilon=0.97$ for the latter case where we have $N=10^6$ players.}
\end{figure}

\begin{figure*}
\centerline{\epsfig{file=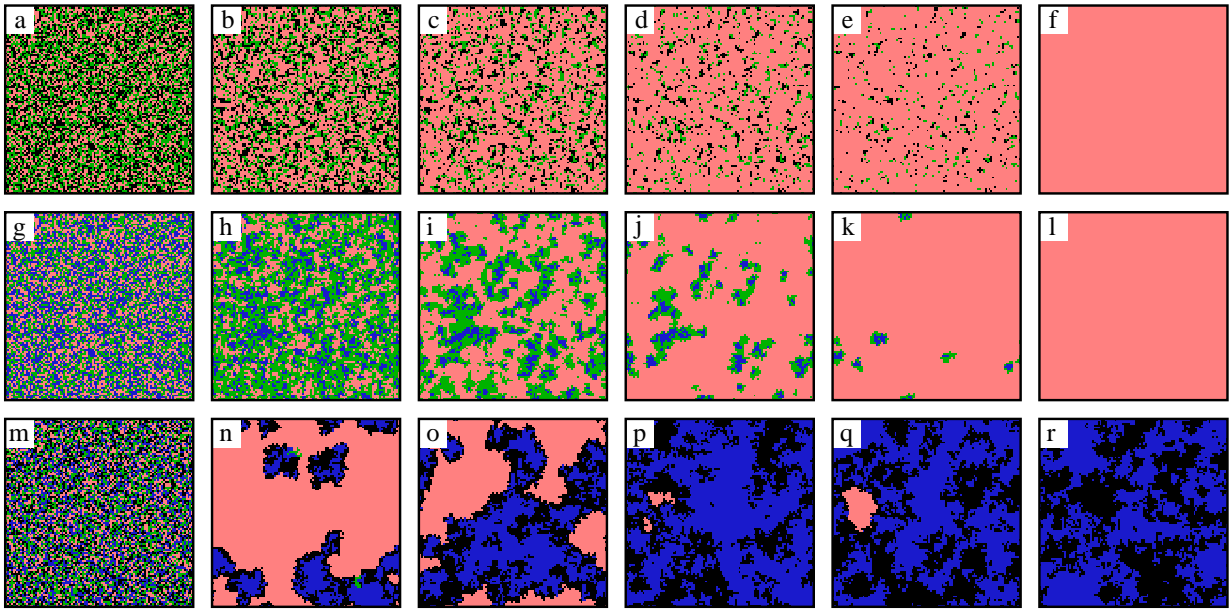,width=16.5cm}}
\caption{\label{snapshot} (color online) Spatial pattern formation explains the evolutionary advantage of alliance with excluder strategy. Every row illustrates different evolutionary trajectory depending on which strategies are present at the beginning. In panels~(a) to (f) only green $C$ (light grey) and black $P$ strategies fight against red (middle grey) defectors. In panels~(g) to (l) only $C$ and blue (dark grey) $E$ strategies compete with $D$ players for space. In both mentioned cases defection prevails because neither punishment nor exclusion alone is able to compete efficiently against defection. However, when they both $P$ and $E$ are present, shown in panels (m) to (r), then they can form a powerful alliance which can crowed out defectors. Parameters are $r=2, \beta=0.6, \gamma=0.01, \epsilon=2.4$, and $L=100$ for all three cases. Note that the triumph of defection is very fast, takes only 100 $MCS$s for the first two rows, but the spread of $P+E$ alliance is relatively slow and 4000 $MCS$s needed to reach a defector-free state.}
\end{figure*}

\subsection{Behaviors in the high $r$ -- low $\gamma$ region}

In the following we turn back to the four-strategy model and focus on the second conceptually different parameter region where the synergy factor is still high enough to ensure coexistence with defectors but the cost of punishment is small enough to reveal the advantage of $P$ strategy. To illustrate the typical behavior in this region we chose the representative parameter values $r=3.8, \beta=0.2$, and $\gamma=0.01$, where in the absence of exclusion $P$ players would crowed out $C$ players. In the absence of the latter strategy punisher players can control defector more efficiently, which provides a significantly higher cooperation level comparing to the $C+D$ solution \cite{helbing_njp10}. Interestingly, here the outcomes of evolution is conceptually similar to the behavior we observed for the two-strategy $D-E$ model. Namely, either $D$ or $E$ players die out during the evolution and the final destination is highly ambiguous especially at small system sizes. Our results are summarized in Fig.~\ref{lCP} where we plotted the probability of defector's extinction in dependence of cost of exclusion. As we already noted, if defectors die out all other strategies become equivalent because neither $P$ nor $E$ have to bear any additional cost anymore. In the alternative case of evolution $E$ dies out first, after $C$, who is not as efficient against $D$ as $P$ strategy, and finally $D$ and $P$ coexist in the stationary state similar to the behavior we observed for the simplified three-strategy ($D,C,P$) model \cite{helbing_njp10}. 

The serious finite-size dependence of the final outcome has similar origin as we observed for the simplified $D-E$ model. Namely, if the system size is too small then it is unlikely that the sufficiently large island of cooperator strategies emerge during the initial transient, which will keep defectors alive. Figure~\ref{lCP} suggests that the defector-free state is very likely even at very high cost of exclusion. Note that the threshold value in the large size limit is almost 2 times higher than the $\epsilon_c=6.749$ value we obtained when $E$ fighted against $D$ alone at this value of synergy factor. Bearing almost double high cost of exclusion seems impossible at first glance, but here the individual viability of strategy $E$ is based on its common success with punishing strategy.

To emphasize the general importance of the alliance of excluders with other cooperator strategies we monitor how fractions of strategies change in time when the system is started from a random initial state. Initially, when defectors are distributed randomly, cooperator strategies are unable to fight efficiently against defectors. This effect is specially remarkable at small $r$ values when network reciprocity alone cannot provide proper help for cooperators. As a consequence, almost all cooperative players die out, but just a little portion of $E$ and $P$ players survive who could form the necessary large island. When this alliance emerges then it can beat defectors who will die out eventually. It is worth stressing that this effect does not limited to lattice topology but can also be observed when interaction graph is random. This is illustrated in the right panel of Fig.~\ref{monitor}, where a random regular graph is used in which players have still $k=4$ nearest neighbors but their connections are rewired randomly \cite{szabo_jpa04}. In the latter case defectors start growing first but their victor is just temporary because the emerging alliance of $P$ and $E$ strategies will crowed out all $D$ players. 

In the next plot we present some characteristic snapshots of pattern formation which illustrate how alliance with excluder players works. For comparison we also plotted evolutions which were taken at identical parameter values but one of the aliance members was missing. In the first row of Fig.~\ref{snapshot} $C$ and $P$ players try to compete against defectors. In agreement with previous observation for the simplified $(D,C,P)$ model their fight is fruitless because the very low value of $r$ and the moderate value of fine $\beta$ prevent them to survive and the system will terminate into a full defection state. In the middle row  of Fig.~\ref{snapshot} $C$ and $E$ strategies fight against $D$, but they loose again: the value of synergy factor is really small, while the cost of exclusion is significant. Finally, in the last row we allow all cooperator strategies to be present simultaneously. As in the previous cases $C$ dies out very soon due to the small value of $r$, but some surviving $P$ and $E$ players can form a viable alliance and their mixture can gradually crowed out defectors. It is important to stress that the mixture of these cooperative strategies is necessary to beat $D$. Otherwise, when they fight individually, defectors revive. This temporary success of defectors can be observed between panels (p) and (q), where some surviving defectors enter a relatively large homogeneous island of punisher strategies. As a results, a compact $D$ spot starts growing immediately and its propagation is only blocked when its frontier meets with the mixture of $P$ and $E$ players again. After the mixed $P+E$ formation becomes successful and eliminates all defectors.

The above described pattern formation gives a deeper insight why the alliance with excluder strategy can be successful against defection even if both members of the alliance is weaker than $D$ players. When a defector is neighboring with a punisher then the latter could be weak, but the presence of an excluder in the group makes $P$ player successful, because $E$ provides the necessary information to exclude the exploiter. Interestingly, if $D$ is neighboring an excluder then the latter can be weaker due the extremely large exclusion cost which involves the extinction of $E$. Therefore, it is always $P$ (or $C$ if $r$ is large enough, as in the case discussed regarding Fig.~\ref{hCP}) is the one who beats $D$, but the success of the fighter is always based on the support of excluder who is just behind the fighter and provides a competitive payoff for cooperator mates within the group. It is important to stress that similar mechanism cannot be observed in a well-mixed system which is modeled by a mean-field theory. In the latter approach the "hiding" of a vulnerable $E$ player from a direct invasion of a defector is not possible because all group members can be reached with equal probabilities within the group. Consequently, the viability of an $E$ strategy is strongly limited by the individual cost $\epsilon$. In a spatial system, however, the range of strategy interaction is limited which offers a slight symmetry breaking and the chance of $P$ (or $C$ at certain conditions) beats $D$ is higher than the frequency that $D$ beats $E$. This is why strategy $E$ can survive for the benefit of the whole community even if such player has to bear an enormously high individual cost.

\begin{figure}
\centerline{\epsfig{file=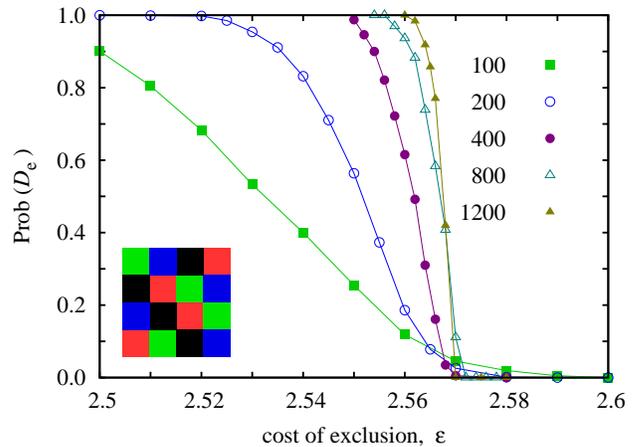,width=9.0cm}}
\caption{\label{lr} (color online) The probability of defector strategy dying out when all 4 strategies are present in the initial state in a patch-like distribution. An example of this kind of starting state is shown in the inset where colors (different range of shades of gray in printed version) mark homogeneous spots of competing strategies. The alternative destination of evolution is the full-$D$ state. System sizes are marked in the legend. Parameters are $r=2, \beta=0.6$, and $\gamma=0.01$. Results were averaged over 5000 independent runs for small sizes while data for the largest size were averaged over 200 runs.}
\end{figure}

\subsection{Behaviors in the low r region}

We close the Results section by presenting representative solutions in the third previously mentioned parameter region where the simplified model predicts qualitatively different behavior \cite{helbing_njp10}. Here  network reciprocity alone is unable to maintain pure cooperation due to the very small synergy factor. As previous studies emphasized, in this case only a very strong punishment can help and when it is fulfilled then the system will terminate into a completely $D$-free state. Otherwise, when the fine is not strong enough, the full-$D$ final destination is inevitable even in the presence of $P$ strategy. The $\beta-\gamma$ dependence of critical fine is illustrated in Ref.~\cite{helbing_ploscb10}. In the present case, when we add strategy $E$, the possible solutions are very similar, but strategy $E$ takes the decisive role of $P$. Namely either $D$ prevails or dies out and leaving a neutral drift among the remaining strategies. But the coexistence of $D$ with other strategies is not possible. More specifically, if the cost of exclusion is low then defection will be eliminated leaving other cooperator strategies alive. Otherwise, when the cost $\epsilon$ is high, defectors will prevail and all other strategies die out. As for higher $r$ value, here we can also observe a serious finite-size effect around the threshold value of $\epsilon$ which has similar origin as explained above. As Figure~\ref{lr} illustrates, both mentioned destinations are possible in a wide interval of $\epsilon$ when the system is not large enough and the solution becomes unambiguous only in the large size limit.

\section{Summary and Conclusion}
\label{summary}

It is our everyday life experience that some members of human societies take extra effort for the community \cite{sober_98, svoboda_13}. From evolutionary viewpoint these group members should be unsuccessful individually and their sacrifice cannot be maintained in the long term. A particular example for such selfless behavior could be an exclusion strategy when a player invests efforts into not only detect defection but also alarms group members about the possible danger. Needless to say, the latter players utilize happily this information and can avoid being exploited by $D$ players. This phenomenon was already studied from evolutionary aspect in well-mixed populations and it was found that the viability of social exclusion is strongly limited by its additional cost $\epsilon$ \cite{sasaki_prsb13, li_k_pre15, li_k_epl16, liu_lj_srep17}. 

The goal of present work was to explore what specific outcomes may be found in structured populations which model the limited number of our interactions more realistically. The first different behavior we can observe in a spatial system is the stable coexistence of defectors and exclusion strategy at specific parameter values, but this solution also requires the presence of pure cooperators. Otherwise, when only defectors meet excluders then only one of them survive, which depends sensitively on the cost of exclusion and the synergy factor $r$. Interestingly, the final outcome is rather ambiguous in a broad cost interval around the threshold value of $\epsilon$ and becomes determined only in the large-size limit. This is a rather unusual behavior in spatial systems where the final destination of evolution is robust and remains largely independent of the initial state even at relatively small system sizes. This behavior can be explained by the fact that the final outcome depends sensitively on whether a critical size of homogeneous excluder domain may emerge, which is a vital condition for their triumph. Evidently, if the system size is too small then the requested domain cannot emerge which leads to a misleading prediction of evolutionary outcome. To clarify this curvature-dependent growth we have made a systematic study and demonstrated that the limit threshold value of exclusion cost obtained from this specific setup agrees with the prediction of large size limit. This analysis revealed that a specific initial state, where competing strategies already form homogeneous clusters could be more efficient to converge faster toward the large size  limit solution.

Our most surprising observation was to reveal that exclusion can be viable and useful to the community even if $E$ excluder strategy has to bear an enormously high individual cost. Such a huge sacrifice would be pointless in a well-mixed population because it would involve the fast extinction of $E$ strategy and the system evolves to an exclusion-free state. In structured populations, however, we can observe the opposite trajectory because defection can be swept out for very high exclusion cost. This behavior is based on a mechanism where exclusion forms an effective defense alliance with other cooperator strategies who do not have to bear the mentioned high cost. When this alliance works then the latter strategy can beat defectors because the simultaneous presence of excluder in the group helps them to avoid being exploited. As we argued, an excluder would be vulnerable when she meets directly with a defector, but this unfortunate meeting happens less frequently than the interaction of $D$ and $P$ (or $C$) players. This is a straightforward consequence of limited range of interactions which is an essential feature of all spatial systems. The latter also explains why conceptually similar behavior may be obtained for other interaction topologies that was also demonstrated for random graph.

Our main results, which cannot be observed in well-mixed populations, highlight that individual efforts should not necessarily be harmful if an effective alliance can be formed where the personal effort is compensated by some protection hence the seemingly trivial individual failure can be avoided via the wellbeing of the whole society. This study also depicts that the viability of an individual strategy should not be evaluated by pair comparison of competing strategies only because spatial systems always offers the chance of a new formation of alliances that can only be discussed when the whole system is on the stage.

\begin{acknowledgments}
This research was supported by the Hungarian National Research Fund (Grant K-120785) and by the National Natural Science Foundation of China (Grant No. 61503062).
\end{acknowledgments}

\end{document}